\documentclass[conference]{IEEEtran}

\IEEEoverridecommandlockouts    

\usepackage{amsmath,amssymb,amsfonts}
\usepackage{algorithmic}
\usepackage{graphicx}
\usepackage{textcomp}
\usepackage{xcolor}

\usepackage{cite}
\usepackage{flushend}
\usepackage{tikz}

\def\BibTeX{{\rm B\kern-.05em{\sc i\kern-.025em b}\kern-.08em
    T\kern-.1667em\lower.7ex\hbox{E}\kern-.125emX}}

\title{\LARGE \bf Modeling of Natural Disasters and Extreme Events in Power System Resilience Enhancement and Evaluation Methods}

\author{Narayan Bhusal, \emph{Student Member, IEEE},  Mukesh Gautam, \emph{Student Member, IEEE},\\ Michael Abdelmalak, \emph{Student Member, IEEE}, and Mohammed Benidris, \emph{Member, IEEE}\\ 
Department of Electrical \& Biomedical Engineering, University of Nevada, Reno, NV\\ 
(emails: \{bhusalnarayan62, mukesh.gautam, mabdelmalak\}@nevada.unr.edu, and mbenidris@unr.edu)}

\begin{document}
\maketitle
\begin{abstract}
The frequency of disruptive and newly emerging threats (e.g. man-made attacks--cyber and physical attacks; extreme natural events--hurricanes, earthquakes, and floods) has escalated dramatically in the last decade. Impacts of these events are very severe ranging from long power outage duration, major power system equipment (e.g. power generation plants, transmission and distribution lines, and substation) destruction, and complete blackout.  Accurate modeling of these events is vitally important as they serve as mathematical tools for the assessment and evaluation of various operations and planning investment strategies to harden power systems against these events. This paper provides a comprehensive and critical review of current practices in modeling of extreme events, system components, and system response for resilience evaluation and enhancement, which is a very important stepping stone toward the development of complete, accurate, and computationally attractive modeling techniques. The paper starts with reviewing existing technologies to model the propagation of extreme events and then discusses the approaches used to model impacts of these events on power system components and system response. This paper also discusses the research gaps and associated challenges, and potential solutions to the limitations of the existing modeling approaches.
\end{abstract}

\section{INTRODUCTION}
\begin{figure}
    \hspace{-2ex}
    \vspace{-1ex}
    \includegraphics[scale=0.34]{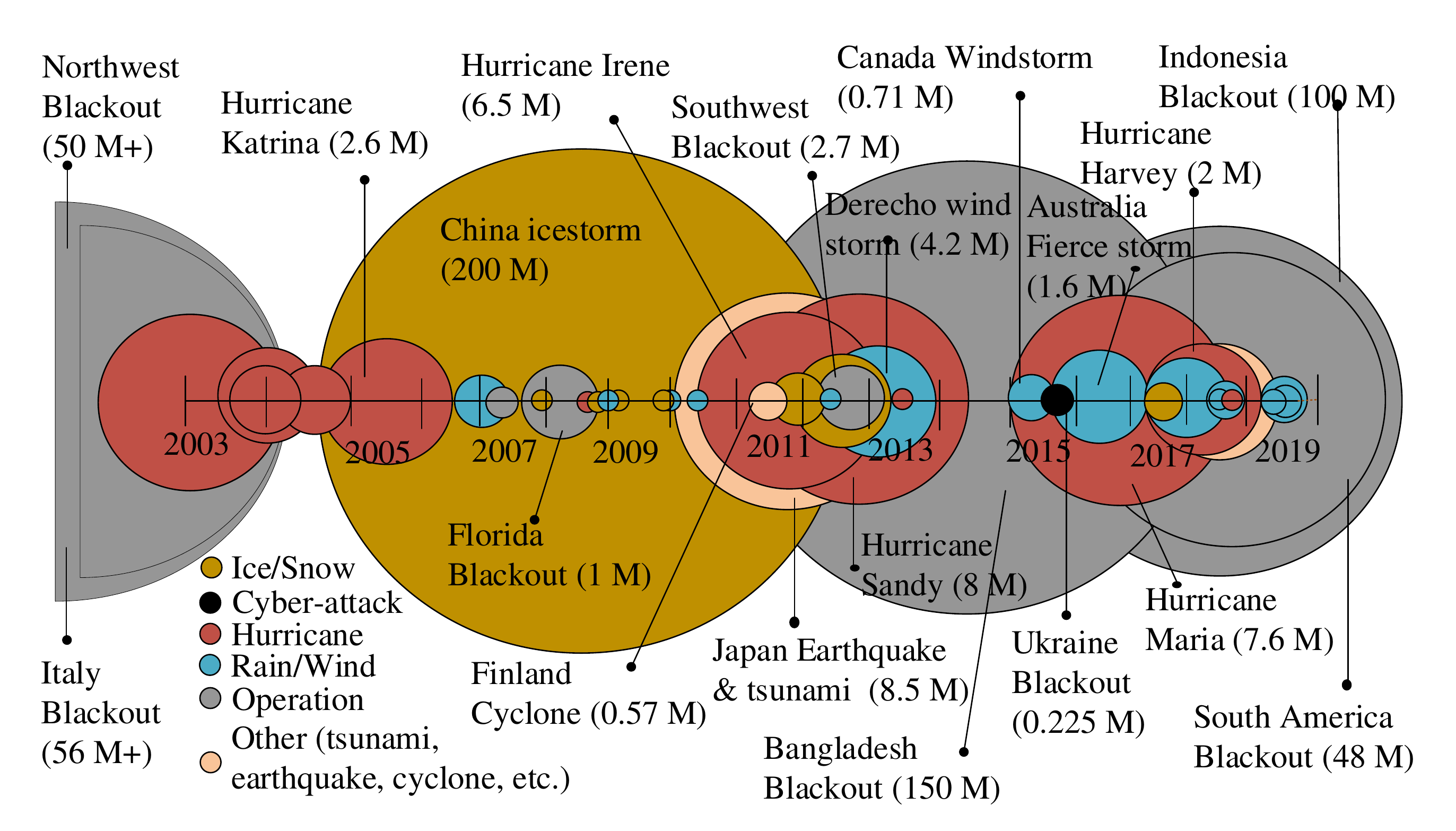}
    \vspace{-3ex}
    \caption{Some of the extreme events, M denotes number of customers affected in million \cite{8966351}}
    \label{fig:extreme_events}
    \vspace{-1.5ex}
\end{figure}

Power system resilience enhancement and evaluation methods have been gaining significant momentum. Although there has been no universally accepted definition for power system resilience, its attributes can be characterized as the ability of power system to ``withstand’’, ``resist’’, and ``recover’’ from disrupting events and ability to proactively respond to potential extreme and newly emerging threats \cite{8966351, SNL1749, JUFRI20191049, PNNL1716,8375946,HUSSAIN201956}. Impacts of extreme events on power system resilience are very severe including: long power outage duration, major power system equipment destruction, cascading failures, and blackouts. Fig. \ref{fig:extreme_events} shows some of the recent extreme events with their impacts in terms power outage to number of customers in millions \cite{8966351}. To evaluate the impacts of extreme events on power system resilience and harden power systems against them, it is important to develop accurate and computationally attractive models for extreme events, component failures due to extreme events, and system responses due to component failures. As several models have been developed and presented in the literature, it is becoming important to discuss their advantages and limitations and their suitability to capture impacts of extreme events on power system resilience. 

Modeling the influence of extreme events and man-made miseries on power system resilience is a very difficult task due to their stochastic and unpredicted nature. Numerous research papers have provided surveys for resilience definitions, metrics, and evaluation and enhancement methods \cite{8966351, JUFRI20191049, 8375946, HUSSAIN201956}. However, they do not provide a comprehensive and critical review on the modeling of extreme events and system and component failures for power system resilience enhancement and evaluation. They mainly either address a specific type of problem (e.g., system recovery), specific type of systems (e.g., distribution systems), or focus on existing definitions, metrics, evaluation methods, and enhancement strategies and compare them with reliability. Further research is needed on modeling of component failures and extreme events for resilience evaluation and enhancement. Our previous work \cite{8966351} provides different modeling aspects for power system resilience; however, it requires further work to completely address power system resilience evaluation from the modeling perspective. Therefore, a review paper that provides a critical and comprehensive review of existing practices, associated challenges, and research gaps with concrete, comprehensive, and constructive recommendations and suggestions for power system resilience is becoming extremely important.

\begin{figure}
    \vspace{-1.5ex}
    \hspace{-3ex}
    \includegraphics[scale=0.64]{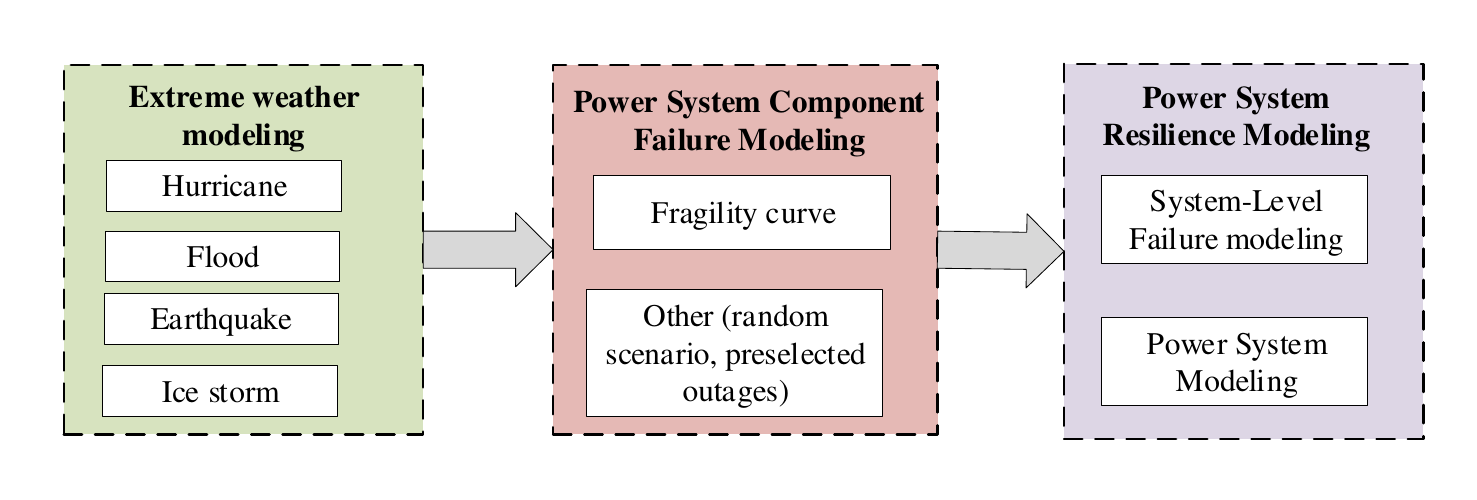}
    \vspace{-5ex}
    \caption{Framework of the Power System Resilience Modeling}
    \vspace{-3ex}
    \label{fig:framework}
\end{figure}

This paper provides a critical and comprehensive review on modeling of power systems for resilience enhancement and evaluation. It starts with reviewing existing modeling technologies to model the evolution and progression of extreme events and then discusses the approaches used to model impacts of these events on power system components and system response.  It also provides discussions on further research needs, associated challenges, and provides potential solutions to the limitations of existing modeling approaches. 
The rest of the paper is organized as follows. Section \ref{Extreme_Event} provides extreme event modeling approaches. Section \ref{failure} discusses critical review on component failure modeling techniques. System modeling for resilience studies is provided in section \ref{system}. Section \ref{future_direction} provides future direction and potential solution for better resilience modeling and concluding remarks are provided in section \ref{Conclusion}. Fig. \ref{fig:framework} provides an overview of power system modeling for resilience studies. 

\section{Modeling of Extreme Weather Events}\label{Extreme_Event}
Natural extreme events such as hurricanes, earthquakes, typhoon, floods, snowstorms, and man-made cyber and physical attack can result in catastrophic consequences to the operation and control of power system \cite{8966351}.  Accurate modeling of these events is very critical to appropriately develop planning and operation strategies. However, due to their stochastic,  spatiotemporal, and unpredicted nature, their accurate modeling is very difficult, exhaustive, and computationally expensive. Each event has a particular impact on the performance of power system. For example, wind storms, ice storms, and hurricanes impact overhead transmission and distribution lines and towers. Floods mostly impact underground cables and power house. Earthquakes may impact underground as well as overhead structures\cite{8299587, 8540318, 8012538}. Most of the literature have used forecasting and historic data from National Weather Service (NWS), Weather Research and Forecasting (WRF), and  National Oceanic and Atmospheric Administration (NOAA) to model the weather-related disasters \cite{7801854, 8606920, 8076912,7885130}. This section provides the overview of the modeling of propagation of extreme events.

\paragraph{Hurricanes}
Hurricane models have been developed based on statistical approaches: probability distribution functions, empirical methods, and sampling approach or through the combination of these approaches \cite{NajafiRavadanegh2019}. The occurrence of hurricanes, direction, angle of propagation, speed, central pressure, speed decaying rate of wind, radius of the wind, etc. need to be incorporated for modeling of hurricane  related disasters \cite{NajafiRavadanegh2019}. A hurricane model has been developed using Poisson distribution function as follows \cite{NajafiRavadanegh2019}. 
\begin{equation}\label{hurricane:piosson}
    P(h)=\frac{\text{exp}({-\lambda})\times \lambda^h}{h!} \text{,}
\end{equation}
where $P$ is the probability distribution function which shows the annual occurrence of hurricanes and $\lambda$ and $h$ are the average number of hurricanes and number of hurricanes per year, respectively. 

The most commonly used hurricane disaster model is HAZUS (hazards US) hurricane model which is developed by the federal emergency management agency (FEMA) \cite{HAZUShurricane}. This model  simulates  hurricane progression based on historic data. The hourly wind profile obtained from organizations like Modern-Era Retrospective Analysis for Research and Application (MERRA) has been used in numerous literature. The intensity, propagation, and time varying impacts of hurricanes have been  determined based on wind speed in \cite{7514755,7036086, 7549241}. Paths of the hurricanes have also been determined based on satellite big data \cite{8540318}, and hourly historic wind profile during hurricane events \cite{7036086, 7434044,7801854}.  

\paragraph{Wind storms}
Extreme wind storms can be modeled using wind extreme simulator which is developed in \cite{windstormmodel}. Authors of \cite{7932183} have used the same simulator to reproduce observed spatial correlation and extreme statistics of adverse winds incorporating the occurrence of wind storms throughout the year. 
\paragraph{Floods}
Floods and estimation of their impacts can be modeled based on the HAZUS flood model developed by FEMA \cite{HAZUSflood}. These models are developed based on data collected from the flood history. In \cite{8076912}, a flood model has been used that is based on rainfall intensities using weather agencies' prediction model.
\paragraph{Ice storms}
The rate of the ice accretion has been modeled based on precipitation rate, wind speed and direction, duration of the ice storm, and liquid water content as follows\cite{8601802}. 
\begin{equation} 
    T_{I}= \frac{n}{\rho_{i}\pi}[(p\rho_{w})^{2}+(3.6 \times v\times w)^{2}]^{1/2} \text{,}
\end{equation}
where $T_{I}$ is the ice thickness; $p$ is the precipitation rate; $n$ represents number of hours of the icy rain; $w$ denotes the liquid water content; $\rho_i$ and $\rho_w$ are, respectively, density of ice and water; and $v$ is the speed of the wind. Ice storms have been modeled based on the forecast of ice storms in \cite{8642947}. Uncertainties associated with the forecasting of ice thickness have also been considered in the model of \cite{8642947}.
\paragraph{Typhoon}
Moving typhoon and its duration has been modeled using Yang Meng wind field model in \cite{8418359}. The equation of typhoon motion has been expressed as follows \cite{MENG1995291}.
\begin{equation}\label{eq:typhoon}
\frac {\partial v}{\partial t} + v \cdot \nabla v = - \frac {1}{\rho }\nabla p - fk \times v + F,\end{equation}
where $p = {p_{\textrm {0}}} + \Delta p\exp \left ({{ - {{\left ({{\frac {{{R_{\textrm {m}}}}}{R}} }\right)}^{B}}} }\right)$ is the mean pressure at the sea level; $v$ is the velocity of wind; $\rho$ is the density of the air; $F$ is frictional force above boundary level; $B$ is pressure constant set as ($0.5~2.5$); $f$ is Coriolis parameter; and $k$ is a fitting parameter. The solution of \eqref{eq:typhoon} provides the velocity of the wind which is required to develop the fragility curve of power system components as discussed in section \ref{failure}. 
\paragraph{Earthquakes}
Similar to the HAZUS model for hurricanes and floods, HAZUS earthquake model has been developed in \cite{HAZUSMR4}. This model uses historical data to model earthquake disasters. Earthquake models usually determine the peak ground acceleration which has been used as input for fragility curves of component failure models. Earthquake models have been developed incorporating intensity of earthquakes, distance between the earthquake center and location of interest, seismic potential, and the type of the ground. A probabilistic earthquake energy transfer model has been proposed based on auto regressive (AR) estimation method in  \cite{8674596}. 
\paragraph{Wildfires}
Wildfire progression has been modeled in \cite{7995099} based on the rate of spread, solar radiation, and radiative heat flux using historical data.

\section{Modeling of Component Failures}\label{failure}
For the evaluation and enhancement of power system resilience, the impact of extreme events leading to the failure of power system can be categorized into component level failure modeling and the system level modeling. The system-level modeling approaches utilize the characteristics and feature of the complete power system for failure modeling, whereas, the component-level modeling approach utilizes probabilistic failure model of a particular component. This section provides a review on component failure models which utilize results obtained from disaster hazard models. 

Failure models of power system components are usually developed based on their probabilistic failure distribution. As it is already mentioned that extreme events are stochastic and unpredictable, power system components have been modeled in the literature using scenario-based methods and fragility functions of the components for specific events. In fragility curve-based methods, failures are modeled based on impacts of event parameters, such as speed of the wind and acceleration during an earthquake, on power system components \cite{8674596, ESPINOZA2016352}.

\subsection{Fragility Curves for Components Failure Modeling} Fragility curves have been used to describe the behavior and vulnerability of system components facing stochastic weather conditions with respect to sequential and regional characteristics \cite{LI2019127}. Fragility curves have been developed based on: a) statistical analysis of large set of historical data; b) experiment; c) analysis of design codes; d) professional judgments; and e) combination of all. In experimental approaches, power system components have been deliberately failed to develop the fragility curves. Analytical approaches have been used when there are no sufficient historical data \cite{7801854, 7932183}. The fragility curve varies according to the event measuring parameters \cite{7036086} and the event severity level \cite{8674596}. For example, failure probabilities of power line towers as a function of speed (fragility function) of wind have been expressed as in \eqref{eq:fragility} \cite{7801854}.
\begin{equation} \label{eq:fragility}
P_{tower}(w) = \left\{ \begin{array}{ll} 0, & \text{if}\; w < {w_{\rm crl}}\\ {P_{tower\_ew}}(w), & \text{if}\; {w_{\rm crl}} \leq w < {w_{\rm cpse}}\\ 1,& \text{if} \; w \geq {w_{\rm cpse}} \end{array} \right.
\end{equation}
where $P_{tower}$ is the probability of tower failure as a function of wind speed,$w$; $ew$ represents the extreme wind; $w_{crl}$ is the speed of the wind above which towers start to experience failure; and $w_{cpse}$ is the wind speed at which tower collapse. This fragility model has been developed by analyzing geometrical and material nonlinearities under a wide range of wind loadings using finite element analysis. 

\begin{figure}
\vspace{-3ex}
    \hspace{-4ex}
    \includegraphics[scale=0.68]{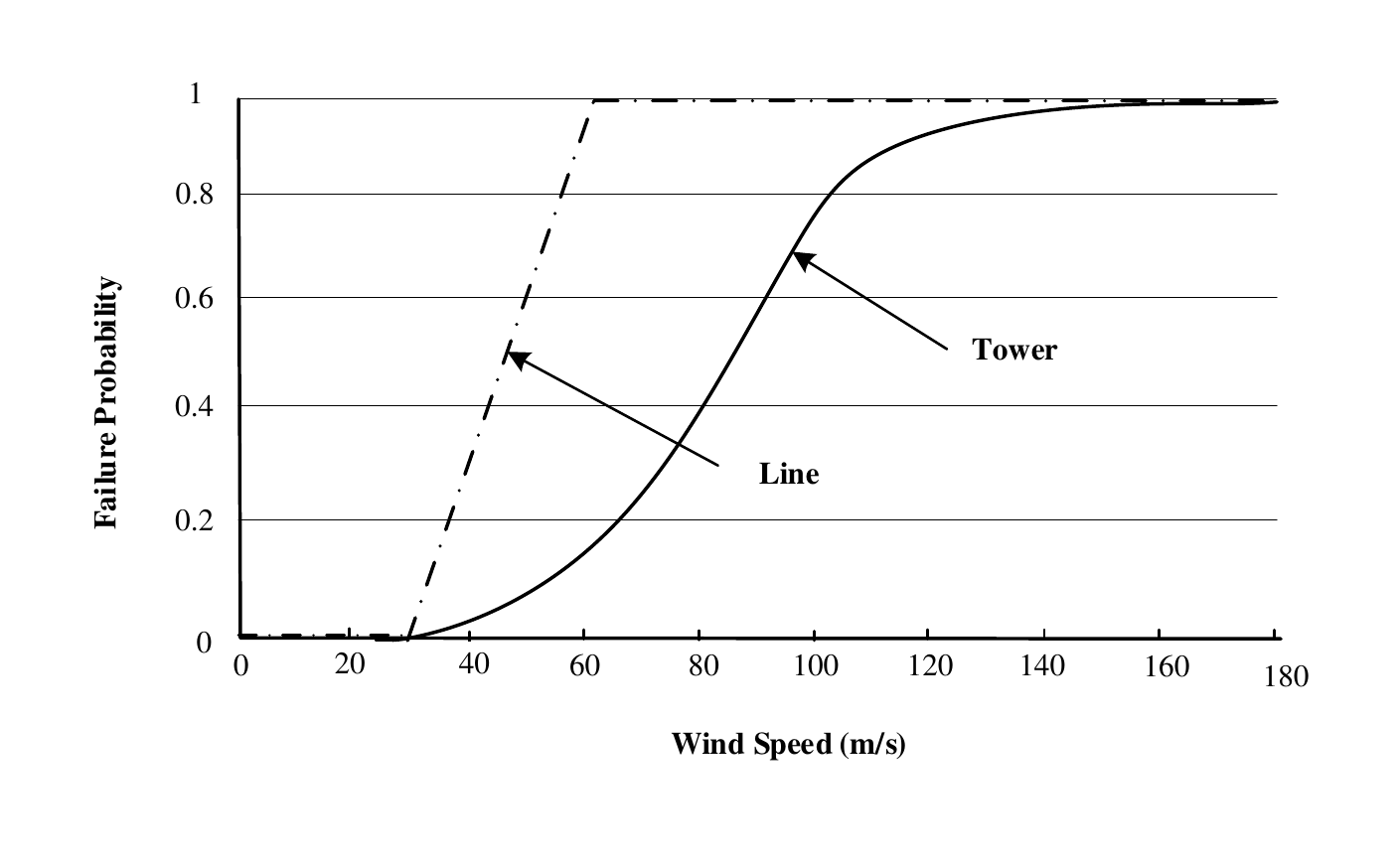}
    \vspace{-7ex}
    \caption{Transmission lines and towers fragility curves \cite{7801854}.}
    \vspace{-2ex}
    \label{fig:fragility}
\end{figure}

The failure model of transmission lines is generally different than that of the pylon from fragility point of view. Therefore, its failure has been modeled separately as in \eqref{eq:fragility_line} \cite{7801854}.  
\begin{equation}\label{eq:fragility_line} 
{P_l}(w) = \left\{ \begin{array}{ll} \overline {{P_{l}}}, &\text{if}\; w < {w_{\rm crl}}\\ {P_{l\_ew}}(w),\ &\text{if}\; {w_{\rm crl}} \leq w < {w_{\rm cpse}}\\ 1, &\text{if}\; w \geq {w_{\rm cse}} \end{array} \right. 
\end{equation}
where $P_l$ is the probability of line failure as a function of wind speed and $\overline{P_{l}}$ is the failure rate at normal weather condition. Fig. \ref{fig:fragility} shows the fragility curves of towers and lines which has been constructed based on the above approach. 

Fragility curve of distribution lines has been developed based on log data of distribution line failures as a function of wind speeds in \cite{8606920}. The fragility curve of power system components due to earthquakes has been developed based on the failure probability of system components as a function of ground acceleration due to earthquakes in \cite{8674596}. As power system is a complex interconnected system, failure of one component may lead to failure of several connected components. Therefore, appropriate modeling of the cascade of these event is vital. Most of fragility curves for various weather disasters for transmission and distribution lines and transmission and distribution towers have been developed based on the approach presented in \cite{7801854, 6616005,7434044, 7932183, HAZUSMR4}. Failure of distribution lines due to hurricanes have been modeled in \cite{JAVANBAKHT2014408} based on static and in-motion gradient wind fields.

\subsection{Other Models} 
Various other approaches other than fragility curve models have been used in the literature for the failure modeling of various power system components in power system resilience studies. Microgrid islanding time and proactive operation strategy has been estimated based on a weather integrated forecasting model in \cite{8012538}. Power distribution systems have been divided into some specific number of territories. An uncertainty modeling technique have been used to determine the number of power outage in the territories \cite{8552410, 7862805}. Transmission line outages have been modeled using the same approach that is proposed in \cite{7885130}. Predefined weather scenarios can also be used to model outages in transmission and distribution lines \cite{8642442, 8642947}. For example, transmission line outages due to an ice storm has been determined based on the forecasted ice thickness in \cite{8642947}. Simulation techniques, such as Monte Carlo simulation (MCS), randomly generate damage scenarios of power system components, which have been also used for modeling of component failures. For example, $10,000$ random scenarios have been generated using MCS to model failures  of power branches in \cite{8586932}. 

\section{System Level Modeling for Resilience Studies}\label{system}
After the overview of disaster hazards model and power system components failure model in section \ref{Extreme_Event} and section \ref{failure}, the overall performance of the system should be evaluated using the proper system-level failure model augmented with the power system models.  

\subsection{System-Level Failure Models}
The system-level failure models utilize historical data of system failure for developing new failure models based on regression and data mining models. The authors of \cite{guikema20101744} have analyzed detailed comparison between regression-based and data mining based models using a statistical validation approach. However, the collection of data is a major challenge in system-level failure modeling.

\paragraph{Regression based models} In \cite{guikema20101744}, two regression based models have been presented, which are a generalized linear model (GLM) and a generalized additive model (GAM). The GLM is a regression-based model consisting of three components: i) a conditional distribution for the count events of the given distribution parameter(s); ii) a link equation that relates explanatory variables and distribution parameter(s); and iii) a regression equation which describes the function of the explanatory variables. On the other hand, the GAM is a regression-based model that account for non-linearities.

\paragraph{Tree-based data mining models} The tree-based data mining approach of modeling has been proposed in \cite{Breiman1984} and is the easiest approach of its kind. The tree-based data mining model uses the recursive binary partitioning of data sets to represent the relationship between response variables of interest and the explanatory variables \cite{guikema20101744}. Two tree-based data mining approaches have been used: classification and regression tree (CART) and Bayesian additive regression tree (BART) \cite{guikema20101744}.

\subsection{Power System Modeling}
The power system model needs to incorporate  system level failure models to develop the complete system resilience model. Numerous models have been developed for power system modeling in resilience studies which vary according to: a) system type; b) resilience improvement techniques; 3) power flow approaches; d) solution approaches; e) technical and operational constraints. As each of the categories plays critical role in the development of power system resilience model, these categories  are further explored for resilience-based studies as follows.

\subsubsection{System Type}
Type of the system is an important aspect that needs to be considered in power system resilience evaluation and enhancement studies. Distribution systems are characterized by radial or weakly-meshed networks; transmission systems are characterized by meshed networks; microgrids are characterized by isolated small radial or weakly-meshed networks (sometimes multiple microgrids are networked together and network of microgrids are formed which increases operational flexibility); and the interdependent systems consist of interconnection of more than one system such as electric power supply system, heat, and gas supply system. As each of these systems has its own unique characteristics, a resilience model should be developed based on their objectives and modeling constraints. For example, power balance, protection device, and unit commitment constraints and power losses need to be considered in the transmission level studies \cite{7434044}. Variable loads, energy storage, distributed energy resources (DERs), and switch status need to considered \cite{7995099} in distribution level studies. 

\subsubsection{Enhancement strategies}
Enhancement strategies are also important power system modeling aspects. Resilience enhancement strategies have been based on utilizing smart grid technologies (e.g., reconfiguration of network, decentralized control, and adaptive restoration); utilization of energy storage, movable energy sources, and various distributed energy resources; and resilience-based crew scheduling for maintenance \cite{EPRI1656, 8966351, SNL1749,PNNL1716,JUFRI20191049,8375946,HUSSAIN201956, PES2020NB}. All of the associated constraints need to be considered in the modeling for resilience studies.

\subsubsection{Power flow models}
Similar to other planning and operational studies, power flow models are essential components in modeling power systems for resilience evaluation and enhancement studies. The main trade-off between existing power flow models are the degree of accuracy and computational time (complexity and simple models with more approximation). LinDistFlow \cite{535709} and DistFlow \cite{25627} methods have been extensively used as power flow models for resilience evaluation of distribution systems as well as microgrids. These models can provide great efficiency and numerically robust solution.  Linear three-phase power flow method developed in \cite{8057667} have also been used for more accurate distribution system resilience modeling. Similarly, for transmission systems, DC, AC, and linearized power flow models have been used. 

\subsubsection{Solution algorithms}
Several solution algorithms have been implemented for the modeling of various resilience evaluation and enhancement strategies. This can be broadly categorized into deterministic, stochastic, and population based methods. Deterministic algorithms include linear programming, mixed integer programming (linear as well as non-linear); and mixed-integer second-order cone programming. Stochastic mixed integer linear programming and stochastic mixed integer non-linear programming are examples of stochastic approaches. Population based intelligent search methods include genetic algorithm and particle swarm optimization. These algorithms optimize the operation of limited resources to enhance power system resilience.

\subsubsection{Technical and operational constraints}
While developing modeling techniques, various operational and technical constraints need to be satisfied. This includes power flow constraints, generation ramp rate constraints, topology constraints (e.g. distribution system may need to maintain its radial topology), line loading constraints, load curtailment constraints, voltage constraints, and frequency constraints. For interdependent system (power, gas, heat), all the associated constraints of complete system need to be considered while modeling these systems.

\section{Research Gaps, Challenges, and Future Directions}\label{future_direction}
Although modeling of extreme events, component failures, and system response have been under extensive studies and development, some perspectives on their research need further development. These research gaps with potential solutions are presented as follows. 
\begin{enumerate}
\item Most of the existing forecasting methods are developed based on several approximation and assumption which compromise their accuracy. Also, meteorological data used for forecasting mostly rely on historical data-set developed from the propagation of a single event in a specific geographical location. Moreover, these data have been assumed fully reliable without accurately modeling the noise, communication, calibration errors, and other various uncertainties. Big data analytics and deep learning methods could be useful to develop better weather forecasting models.

\item Existing fragility curves cannot capture  spatiotemporal effect of extreme weather events as they can not provide accurate realization and propagation of these events and their impacts on the failure of power grid components. Integrating scenario-based simulation methods with more accurate fragility curves could provide a means to develop holistic and accurate failure models.

\item  The correlation between interdependent critical infrastructure (power supply, water supply, road structure) system should be extensively studied to capture the impact of the failure of each system element on the other connected system elements. 
\item The dynamic behavior of renewable energy sources, battery energy storage system, mobile emergency resources is usually neglected in microgrid islanding and formation approaches because of the associated uncertainties such as weather-related variabilities. Monte Carlo simulation could be coordinated with other analytical approaches to appropriately model their uncertainties. 
\item Most of the literature have only considered 24 hours of scheduling horizon. Short scheduling horizon has negative impacts on power system, especially power systems with high penetration of distributed energy resources. This could result in load shedding to critical loads in following days while providing power to non-critical loads in the current day. Long scheduling horizon is required to better utilize the limited resources during disasters; parallel optimization approach can be used for these multi-stage long scheduling horizon simulation problem.
\item While developing models for distribution systems, radial topologies as well weakly-meshed networks should be properly considered as both of these topologies exist in practical distribution systems.
\item Demand response programs should be developed to utilize the concept of demand-side management during the disasters.
\item Lack of interconnection standards (IEEE-1547 requires customer-owned energy resource to disconnect from the grid during restoration because of safety and power quality concern) and appropriate compensation scheme is holding the use of customer-owned energy resources for service restoration during disasters. Appropriate policies and standard need to be developed from respective institutions.
\item The availability of perfect information could be very difficult during the disasters therefore. Therefore, modeling approaches should be developed for incomplete and imperfect information.
\end{enumerate}

\section{Conclusion}\label{Conclusion}
This paper has provided a comprehensive and critical review of current practices of extreme events modeling from the perspective of power system resilience evaluation and enhancement. Complete power system resilience modeling approaches were also provided which include modeling of system components as well power system response during extreme events. Fragility curves and random scenario-based simulation approaches have been the main modeling approaches to model failure of power system components due to extreme events. Finally, this paper discussed the research gaps, associated challenges, and potential solutions to the limitations of the existing modeling approaches. Accurate and mathematically attractive models for extreme events, system components, and system response are still needed for power system resilience evaluation and enhancement.
 \section*{Acknowledgment} 
 This work was supported by the U.S. National Science Foundation (NSF) under Grant NSF 1847578.
\bibliographystyle{IEEEtran}
\bibliography{References.bib}
\end{document}